\documentclass{llncs}
\usepackage{multirow}
\usepackage{article}
\usepackage{longtable}
\makeatletter
\newtoks\fintableau
\let\fintableau\@arraycr
\makeatother
\begin{document}


\title{Real-Time Specification Patterns and
  Tools\thanks{\footnotesize{This work was partially supported by the
      JU Artemisia project CESAR and the FNRAE project Quarteft}}}
\author{Nouha Abid\inst{1,2} \and Silvano {Dal~Zilio}\inst{1,2} \and
  Didier {Le~Botlan}\inst{1,2}}
\institute{CNRS, LAAS, 7 avenue du colonel Roche, F-31400 Toulouse
  France \and Univ de Toulouse, LAAS, F-31400 Toulouse, France}
\maketitle
\setcounter{footnote}{0}
     
\begin{abstract}
  An issue limiting the adoption of model checking technologies by the
  industry is the ability, for non-experts, to express their
  requirements using the property languages supported by verification
  tools. This has motivated the definition of dedicated assertion
  languages for expressing temporal properties at a higher
  level. However, only a limited number of these formalisms support
  the definition of timing constraints. In this paper, we propose a
  set of specification patterns that can be used to express real-time
  requirements commonly found in the design of reactive systems. We
  also provide an integrated model checking tool chain for the
  verification of timed requirements on TTS, an extension of Timed
  Petri Nets with data variables and priorities.
\end{abstract}

\section{Introduction}
\label{sec:introduction}

An issue limiting the adoption of model checking technologies by the
industry is the difficulty, for non-experts, to express their
requirements using the specification languages supported by the
verification tools. Indeed, there is often a significant gap between
the boilerplates used in requirements statements and the low-level
formalisms used by model checking tools; the latter usually relying on
temporal logic. This limitation has motivated the definition of
dedicated assertion languages for expressing properties at a higher
level (see Section~\ref{sec:related-work}). However, only a limited
number of assertion languages support the definition of timing
constraints and even fewer are associated to an automatic verification
tool, such as a model checker.

In this paper, we propose a set of real-time specification patterns
aimed at the verification of reactive systems with hard real-time
constraints. Our main objective is to propose an alternative to timed
extensions of temporal logic during model checking. Our patterns are
designed to express general timing constraints commonly found in the
analysis of real-time systems (such as compliance to deadlines; event
duration; bounds on the worst-case traversal time; etc.). They are
also designed to be simple in terms of both clarity and computational
complexity. In particular, each pattern should correspond to a
decidable model checking problem.

Our patterns can be viewed as a real-time extension of Dwyer's
specification patterns~\cite{ppsfsv1999}. In his seminal work, Dwyer
shows through a study of 500 specification examples that 80\% of the
temporal requirements can be covered by a small number of ``pattern
formulas''. We follow a similar philosophy and define a list of
patterns that takes into account timing constraints. At the syntactic
level, this is mostly obtained by extending Dwyer's patterns with two
kind of \emph{timing modifiers}: (1) $P$ \code{within} $I$, which
states that the delay between two events declared in the pattern $P$
must fit in the time interval $I$; and (2) $P$ \code{lasting} $D$,
which states that a given condition in $P$ must hold for at least
duration $D$. For example, we define a pattern {\code{Present} $A$
  \code{after} $B$ \code{within} $]0, 4]$} to express that the event
$A$ must occur within 4 units of time (u.t.) of the first occurrence
of event $B$, if any, and not simultaneously with it. Although
seemingly innocuous, the addition of these two modifiers has a great
impact on the semantics of patterns and on the verification techniques
that are involved.

Our second contribution is an integrated model checking tool chain that
can be used to check timed requirements. We provide a compiler for
Fiacre~\cite{filfmvte2008}, a formal modelling language for real-time
systems, that we extended to support the declaration of real-time
patterns. In our tool chain, Fiacre is used to
express the model of the system while verification activities
ultimately relies on Tina~\cite{tina}, the TIme Petri Net
Analyzer. This tool chain provides a reference implementation for our
patterns when the systems can be modeled using an extension of Time
Petri Nets with data variables and priorities that we call a TTS (see
Sect.~\ref{sec/tts}). This is not a toy example; Fiacre is the
intermediate language used for model verification in
Topcased~\cite{ttptosfcasd2006}, an Eclipse-based toolkit for system
engineering, where it is used as the target of model transformation
engines for various high-level modelling languages, such as SDL or
AADL~\cite{aadl2fcr}. In each of these transformations, we have been
able to use our specification patterns as an intermediate format
between high-level requirements (expressed on the high-level models)
and the low-level input languages supported by the model checkers in
Tina.

The rest of the paper is organized as follows. In the next section, we
define technical notations necessary to define the semantics of
patterns. Section~\ref{sec4:RPC} gives our catalog of real-time
patterns. For each pattern, we give a simple definition in natural
language as well as an unambiguous, formal definition based on two
different approaches. Before concluding, we review the results of
experiments that have been performed using our verification tool chain
in Sect.~\ref{sec:use-cases-exper}.

\section{Technical Background}
\label{back}

Since patterns are used to express timing and behavioral constraints
on the execution of a system, we base the semantics of patterns on the
notion of \emph{timed traces}, which are sequences mixing events and
time delays, (see Def.~\ref{def/timed-trace} below). We use a dense
time model, meaning that we consider rational time delays and work
both with strict and non-strict time bounds. 

The semantics of a pattern will be expressed as the set of all timed
traces where the pattern holds. We use two different approaches to
define set of traces: (1) Time Transition Systems (TTS), whose
semantics relies on timed traces; and (2) a timed extensions of Linear
Temporal Logic, called MTL. In our verification tool chain, both Fiacre
and specification patterns are compiled into TTS.

\subsection{Time Transition Systems and Timed Traces}
\label{sec/tts}

\begin{figure}[h]
\centerline{
\begin{tikzpicture}[node distance=6ex and 9.5ex, label distance=-0.5ex]
\node[place, label=below:Idle, tokens=1] (s0) {} ;
\node[transition, above=of s0, label=left:{Shutdown},label=above:{\preact{$\neg (\lbl{req}_1 \vee \lbl{req}_2$)}}] (toff) {}
  edge[pre] (s0) ;
\node[vtransition, right=of s0, label=below:{Ventil.}, label=above:{$[6;6]$}] (t1) {}
  edge[post] (s0) ;
\node[place, label=below:Refresh, right=of t1]  (s1) {}
  edge[post] (t1) ;
\path(s1) +(15:15ex) node[vtransition, label={above:\begin{tabular}{c}Close$_1$\\\postact{req$_1$ := false}\end{tabular}}, label={[label distance=-1.15ex] below right:$[4;4]$}] (t2) {} 
  edge[post] (s1) ;
\node[place, label=right:D$_1$isOpen, right=11ex of t2] (s2) {}
  edge[post] (t2) ;
\path(s1) +(-15:15ex) node[vtransition, label={below:\begin{tabular}{c}\postact{req$_2$ := false}\\Close$_2$\end{tabular}}, label={[label distance=-1.15ex] above right:$[4;4]$}] (t3) {} 
  edge[post] (s1) ;
\node[place, label=right:D$_2$isOpen, right=11ex of t3] (s3) {}
  edge[post] (t3) ;
\node[vtransition, above=9.5ex of s1, label={above:\begin{tabular}{c}Open$_1$\\\preact{$\lbl{req}_1$}\end{tabular}}, label={below:$[0,0]$}] (t4) {}
  edge[post, out=0, in=90] (s2)
  edge[pre, out=180, in=45]  (s0) ;
\node[vtransition, below=9.5ex of s1, label={below:\begin{tabular}{c}\preact{$\lbl{req}_2$}\\Open$_2$\end{tabular}}, label={above:$[0,0]$}] (t5) {}
  edge[post, out=0, in=-90] (s3)
  edge[pre, out=180, in=-45]  (s0) ;
\node[transition, left=of s0, label=left:{Button$_1$}, label=below:{\postact{req$_1$ := true}}, label=above:{\preact{$\neg \lbl{req}_1$}}] (b1) {} ;
\node[transition, below=9ex of b1, label=left:{Button$_2$}, label=below:{\postact{req$_2$ := true}}, label=above:{\preact{$\neg \lbl{req}_2$}}] (b2) {} ;
\draw[prio] (t5) to[ out=150, in=-150] (t4) ;
\end{tikzpicture}
}
\caption{A TTS model of an airlock system}
\label{fig/dble-TTS}
\end{figure}
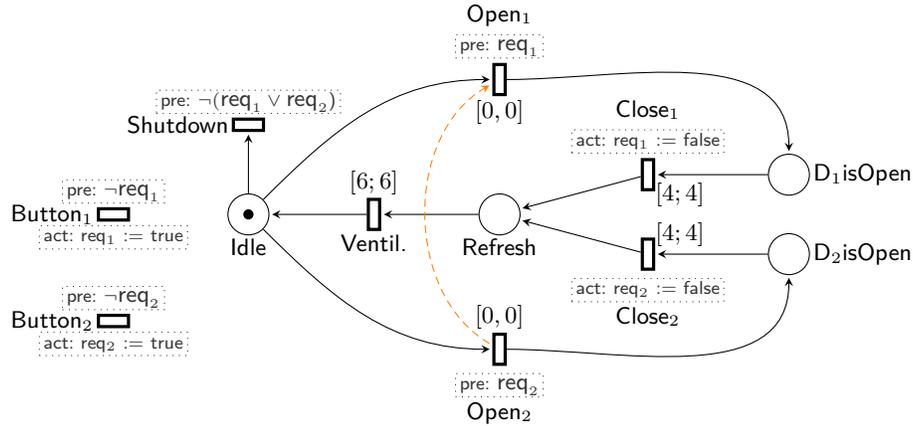

Time Transition Systems (TTS) are a generalization of Time Petri
Nets~\cite{merlin} with priorities and data variables. We describe the
semantics of TTS using a simple example. Figure~\ref{fig/dble-TTS}
gives a model for a simple airlock system consisting of two doors
($D_1$ and $D_2$) and two buttons. At any time, at most one door can
be open. This constraint is modeled by the fact that at most one of
the places \lbl{D$_1$isOpen} and \lbl{D$_2$isOpen} may have a token.
Additionally, an open door is automatically closed after exactly 4
units of time (u.t.), followed by a ventilation procedure that lasts
6~u.t. This behavior is modeled by adding timing constraints on the
transitions \lbl{Open$_1$}, \lbl{Open$_2$} and \lbl{Ventil}. Moreover,
requests to open the door~$D_2$ have higher priority than requests to
open~$D_1$. This is modeled using a priority (dashed arrow) from the
transition \lbl{Open$_2$} to \lbl{Open$_1$}. A shutdown command can be
triggered if no request is pending.

To understand the model, the reader may, at first, ignore side
conditions and side effects (the \code{pre} and \code{act} expressions
inside dotted rectangles). In this case, a TTS is a standard Time
Petri Net, where circles are places and rectangles are transitions.  A
transition is enabled if there are enough tokens in its input
places. A time interval, such as $I = [d_1; d_2[$, indicates that the
corresponding transition must be fired if it has been enabled for $d$
units of time with $d \in I$. As a consequence, a transition
associated to the time interval $[0;0]$ must fire as soon as it is
enabled.  Our model includes two boolean variables, \lbl{req$_1$} and
\lbl{req$_2$}, indicating whether a request to open door $D_1$
(resp. $D_2$) is pending. Those variables are read by pre-conditions
on transitions \lbl{Open$_i$}, \lbl{Button$_i$}, and \lbl{Shutdown}
and are modified by post-actions on transitions \lbl{Button$_i$} and
\lbl{Close$_i$}. For instance, the pre-condition $\neg \lbl{req}_2$ on
\lbl{Button}$_2$ is used to disable the transition when the door is
already open. This implies that pressing the button while the door is
open has no further effect.

We introduce some basic notations used in the remainder of the
paper. (A complete, formal description of the TTS semantics can be
found in~\cite{VRTSPTTS}.)  Like with Petri Nets, the state of a TTS
depends on its marking, $\M$, that is the number of tokens in each
place. We write $\Markings$ the set of markings. Since we manipulate
values, the state of a TTS also depends on its \emph{store}, that is a
mapping from variable names to their respective values.  We use the
symbol $s$ for a store and write $\Stores$ for the set of
stores. Finally, we use the symbol $t$ for a transition and $\T$ for
the set of transitions of a TTS.  The behavior of a TTS is defined by
the set of all its (timed) traces.  In this particular case, a trace
will contain information about fired transitions
(e.g. \lbl{Open$_1$}), markings, the value of variables, and the
passing of time. Formally, we define an event $\event$ as a triple
$(t, \M, s)$ recording the marking and store {immediately after} the
transition $t$ has been fired. We denote $\Events$ the set $\T
{\times} \Markings {\times} \S$ of events.  The set of non-negative
rational numbers is written $\rationalp$.
\begin{definition}[Timed trace]
\label{def/timed-trace}
A timed trace $\s$ is a possibly infinite sequence of events $\event
\in \Events$ and durations $d(\delay)$ with $\delay \in
\rationalp$. Formally, $\s$ is a partial mapping from $\nat$ to
$\Events^* = \Events \cup \set{d(\delay) \mid \delay \in \rationalp}$
such that $\s(i)$ is defined whenever $\s(j)$ is defined and $i \leq
j$.
\end{definition}
Given a finite trace $\s$ and a---possibly infinite---trace $\s'$, we
denote $\s \s'$ the \emph{concatenation} of $\s$ and $\s'$. This
operation is associative. The semantics of patterns will be defined as
a set of timed traces. Given a real-time pattern $P$, we say that a
TTS $T$ satisfies the requirement $P$ if all the traces of $T$ hold
for $P$.

\subsection{Metric Temporal Logic and Formulas over Traces}

Metric Temporal Logic (MTL)~\cite{SRPMTL} is an extension of LTL where
temporal modalities can be constrained by a time interval. For
instance, the MTL formula $A \until_{[1,3[} B$ states that in every
execution of the system (in every trace), the event $B$ must occur at
a time $t_0 \in [1,3[$ and that $A$ holds everywhere in the interval
$[0,t_0[$. In the following, we will also use a weak version of the
``until modality'', denoted $A \W B$, that does not require $B$ to
eventually occur. We refer the reader to~\cite{ow05} for a
presentation of the logic and a discussion on the decidability of
various fragments.

An advantage of using MTL is that it provides a sound and unambiguous
framework for defining the meaning of patterns. Nonetheless, this
partially defeats one of the original goal of patterns, that is to
circumvent the use of temporal logic in the first place. For this
reason, we propose an alternative way for defining the semantics of
patterns that relies on first-order formulas over traces. For
instance, when referring to a timed trace $\sigma$ and an event $A$,
we can define the ``scope'' $\sigma$ \code{after} $A$--that determines
the part of $\sigma$ located after the first occurrence of $A$--as the
trace $\sigma_2$ such that $\exists \sigma_1 . \sigma = \sigma_1 A
\sigma_2 \wedge A \notin \sigma_1$. We believe that this second
approach may ease the work of engineers that are not trained with
formal verification techniques. Our experience shows that being able
to confront different definitions for the same pattern, using
contrasting approaches, is useful for teaching patterns.

\subsection{Model checking, Observers and TTS}

We have designed our patterns so that checking whether a system
satisfies a requirement is a decidable problem. We assume here that we
work on discrete models (with a continuous time semantics), such as
timed automata or time Petri Nets, and not on hybrid models. Since the
model checking problem for MTL is undecidable~\cite{ow05}, it is not
enough to simply translate each pattern into a MTL formula to check
whether a TTS satisfies a pattern. This situation can be somehow
alleviated. For instance, the problem is decidable if we disallow
simultaneous events in the system and if we disallow punctual timing
constraints, of the form $[d, d]$. Still, while we may rely on timed
temporal logics as a way to define the semantics of patterns, it is
problematic to have to limit ourselves to a decidable fragment of a
particular logic--which may be too restrictive--or to rely on multiple
real-time model checking algorithms--that all have a very high
complexity in practice.

To solve this problem, we propose to rely on \emph{observers} in order
to reduce the verification of timed patterns to the verification of
LTL formulas. We provide for each pattern, $P$, a pair $(T_P, \phi_P)$
of a TTS model and a LTL formula such that, for any TTS model $T$, we
have that $T$ satisfies $P$ if and only if $T \otimes T_P$ (the
composition of the two models $T$ and $T_P$) satisfies $\phi_P$.
The idea is not to provide a generic way of obtaining the observer
from a formal definition of the pattern. Rather, we seek, for each
pattern, to come up with the best possible observer in practice. To
this end, using our tool chain, we have compared the complexity of
different implementations on a fixed set of representative examples
and for a specific set of properties
and kept the best candidates.

\section{A Catalog of Real-Time Patterns}
\label{sec4:RPC}

We describe our patterns using a hierarchical classification borrowed
from Dwyer~\cite{ppsfsv1999} but adding the notion of ``timing
modifiers''. Patterns are built from five categories, listed below, or
from the composition of several patterns (see
Sect.~\ref{realtimepatternsC}):
\begin{itemize}
\item \textbf{Existence Patterns (Present):} for conditions that must
  eventually occur;
\item \textbf{Absence Patterns (Absent):} for conditions that should
  not occur;
\item \textbf{Universality Patterns :} for conditions that must occur
  throughout the whole execution;
\item \textbf{Response Patterns (Response):} for (trigger) conditions
  that must always be followed by a given (response) condition;
\item \textbf{Precedence Patterns :} for (signal) conditions that must
  always be preceded by a given (trigger) condition.
\end{itemize}
In each class, generic patterns may be specialized using one of five
\emph{scope modifiers} that limit the range of the execution trace
over which the pattern must hold:
\begin{itemize}
\item \textbf{Global :} the default scope modifier, that does not
  limit the range of the pattern. The pattern must hold over the whole
  timed trace;
\item \textbf{Before R :} limit the pattern to the beginning of a time
  trace, up to the first occurrence of R;
\item \textbf{After Q :} limit the pattern to the events following the
  first R;
\item \textbf{Between Q and R :} limit the pattern to the events
  occurring between an event Q and the following occurrence of an
  event R;
\item \textbf{After Q Until R :} similar to the previous scope
  modifier, except that we do not require that R must necessarily
  occur after a Q.
\end{itemize}
Finally, timed patterns are obtained using one of two possible kind of
\emph{timing modifiers} that limit the possible dates of events
referred in the pattern:
\begin{itemize}
\item \textbf{Within $I$ :} to constraint the delay between two given
  events to belong to the time interval $I$;
\item \textbf{Lasting $D$ :} to constraint the length of time during
  which a given condition holds (without interruption) to be greater
  than $D$.
\end{itemize}

When defining patterns, the symbols $A, B, \dots$ stand for predicates
on events $\omega \in \Omega$ such as $\code{Open}_2 \vee
\code{req}_2$. In the definition of observers, a predicate $A$ is
interpreted as the set of transitions of the system that match
$A$. Due to the somewhat large number of possible alternatives, we
restrict this catalog to the most important presence, absence and
response patterns. Patterns that are not described here can be found
in a long version of this paper~\cite{FRP11}.

For each pattern, we give its denotational interpretation based on
First-Order formulas over Timed Traces (denoted FOTT in the following)
and a logical definition based on MTL. We provide also the observer
and the LTL formula that should be combined with the system in order
to check the validity of the pattern. We define some conventions on
observers. In the following, \lbl{Error}, \lbl{Start}, \dots are
transitions that belong to the observer, whereas \lbl{E}$_1$
(resp. \lbl{E}$_2$) represents all transitions of the system that
match predicate $A$ (resp. $B$). We also use the symbol $I$ as a
shorthand for the time interval $[d_1, d_2]$. The observers for the
pattern obtained with other time intervals--such as $]d_1, d_2]$,
$]d_1, +\infty[$, or in the case $d_1 = d_2$--are essentially the
same, except for some priorities between transitions that may change.
By convention, the boolean variables used in the definition of an
observers are initially set to false.


\subsection{Existence patterns}
\label{realtimepatternsP}

An existence pattern is used to express that, in every trace of the
system, some events must occur.
\smallskip

\fichepattern
{\code{Present} $A$ \code{after} $B$ \code{within} $I$}
{Predicate $A$ must hold
between $d_1$ and $d_2$ units of time (u.t) after the first occurrence
of $B$. The pattern is also satisfied if $B$ never holds.
}
{}
{\pop{present} \code{Ventil.} \pop{after} $\code{Open}_1 \vee \code{Open}_2$ \pop{within} $[0,10]$}
{(\neg B) \W (B \wedge \mathit{True} \until_{I} A)}
{\forall \s_1,\s_2 \such (\s = \s_1 B \s_2 \wedge B \notin \s_1)
 \imply 
\exists \s_3, \s_4 \such \s_2 = \s_3 A \s_4 \wedge \Delta(\s_3) \in I
}
{\begin{tikzpicture}[baseline,node distance=9ex and 17ex, label distance=-0.4ex]
     \node[vtransition, label=below:{\postact{foundB := true}}, label=left:E$_2$] (t1) {};
     \node[vtransition, right=of t1, label=left:Start, label=above:{\preact{foundB}},label=below:{\postact{flag := true}}, label=right:{$[d_1,d_1]$}] (t3) {} ;
     \node[vtransition, right=of t3, label=below:{\postact{\begin{tabular}[t]{l}if flag then\\foundA := true\end{tabular}}}, label=left:E$_1$] (t2) {};
     \node[vtransition, right=of t2, label=left:Error, label=above:{\preact{foundB $\wedge \neg$ foundA}},label=right:{$[d_2,d_2]$}] (t4) {} ;
    \draw[prio] (t2.south west) -- (t4.south east) ;
    \draw[prio] (t3.south east) -- (t2.south west) ;
      
    \end{tikzpicture}
}
{[]\neg \lbl{Error}}

\noindent  In this observer, transition \code{Error} is conditioned by the value
of the shared boolean variables \code{foundA} and \code{foundB}.
Variable \code{foundB} is set to true after transition \code{$E_2$}
and transition \code{Error} is enabled only if the predicate
\code{foundB} $\wedge \neg\,$ \code{foundA} is true. Transition
\code{Start} is fired $d_1$ u.t after an occurrence of \code{$E_2$}
(because it is enabled when \code{foundB} is true). Then, after the
first occurrence of \code{$E_1$} and if \code{flag} is true,
\code{foundA} is set to true. This captures the first occurrence of
\code{$E_1$} after \code{Start} has been fired. After $d_2$ u.t., in
the absence \code{$E_1$}, transition \code{Error} is fired. Therefore,
the verification of the pattern boils down to checking if the event
\code{Error} is reachable. The priority (dashed arrows) between
\code{Start}, \code{Error}, and \code{$E_1$} is here necessary to
ensure that occurrences of \code{$E_1$} at precisely the date $d_1$ or
$d_2$ are taken in account.
\smallskip

\fichepattern
{\code{Present first} $A$ \code{before} $B$ \code{within} $I$}
{The first occurrence of predicate $A$ holds between $d_1$ and $d_2$ u.t. before the first occurrence of~$B$.
The pattern is also satisfied if $B$ never holds. (The difference with
\code{Present} $B$ \code{after} $A$ \code{within} $I$ 
 is that $B$ should not occur before the first $A$.) 
}{}
{\pop{present first} $\code{Open}_1 \vee \code{Open}_2$ \pop{before} \code{Ventil.} \pop{within} $[0,10]$}
{(\ltlexist B) \imply (\,(\neg A \wedge \neg B) \until (A \wedge \neg B \wedge (\neg B \until_{I} B))\,)}
{%
\forall \s_1, \s_2 \such (\s = \s_1 B \s_2 \wedge B \notin \s_1)
 \imply 
\exists \s_3, \s_4 \such \s_1 = \s_3 A \s_4 
 \wedge A \notin \s_3 \wedge \Delta(\s_4)\in{I}
}
{
\begin{tikzpicture}[baseline,node distance=9ex and 17ex, label distance=-0.4ex]
     \node[vtransition, label=below:{\postact{foundA := true}}, label=left:E$_1$] (t2) {};
     \node[vtransition, right=of t2, label=left:Start, label=above:{\preact{foundA}},label=below:{\postact{flag := true}}, label=right:{$[d_1,d_1]$}] (t3) {} ;
     \node[vtransition, right=of t3, label=below:{\postact{foundB := true}}, label=left:E$_2$] (t1) {};
     \node[vtransition, right=of t1, label=left:Error, label=above:{\preact{foundA $\wedge \neg$ foundB}},label=right:{$[d_2,d_2]$}] (t4) {} ;
     \draw[prio] (t3.south east) -- (t1.south west) ;
     \draw[prio] (t1.south west) -- (t4.south east) ;
  \end{tikzpicture}
}
{(\ltlexist B) \imply \neg \ltlexist (\lbl{Error} \vee (\lbl{foundB} \wedge \neg \lbl{flag}))}

\noindent Like in the previous case, variables \code{foundA} and
\code{foundB} are used to record the occurrence of transitions $E_1$
and $E_2$.  Transition \code{Start} is fired, and variable \code{flag}
is set to true, $d_1$ u.t. after the first \code{$E_1$}. Then
transition \code{Error} is fired only if its precondition---the
predicate \code{foundA} $\wedge \neg\,$ \code{foundB}---is true for
$d_2$ u.t. Therefore transition \code{Error} is fired if and only if
there is an occurrence of \code{$E_2$} before \code{$E_1$} (because
then \code{foundB} is true) or if the first occurrence of \code{$E_2$}
is not within $[d_1,d_2]$ of the first occurrence of \code{$E_1$}.
\smallskip

\fichepattern{\code{Present} $A$ \code{lasting} $D$}
{Starting from the first occurrence when the predicate $A$ holds, it remains true for at least duration $D$.}
{The pattern makes sense only if $A$ is a predicate on states (that
  is, on the marking or store); since transitions are instantaneous,
  they have no duration.}
{\pop{present} \code{Refresh} \pop{lasting} $6$}
{(\neg A) \until (\ltlall_{[0,D]} A)}
{\exists \s_1, \s_2, \s_3 \such \s = \s_1 \s_2 \s_3 \wedge A \notin \s_1 \wedge \Delta(\s_2) \geqslant D \wedge A(\s_2)}
{
\begin{tikzpicture}[baseline,node distance=5ex and 20ex, label distance=-0.5ex]
      \node[vtransition, label=above:{\preact{$A$}}, label=below:{\postact{win := true}}, label=left:{OK}, label=right:{$[D,D]$}] (t2) {};
      \node[vtransition, label=above:{\preact{$A \wedge \neg$ foundA}}, label=below:{\postact{foundA := true}}, label=left:{Poll}, left=of t2] (t1) {};
      \node[vtransition, label=above:{\preact{foundA $\wedge \neg$ win}}, label=left:{Error}, label=right:{$[D,D]$},right=of t2] (t3) {};
      \draw[prio] (t3.south west) -- (t2.south east);
      
     \end{tikzpicture}
}
{\ltlall \neg \lbl{Error}}

\noindent Variable \code{foundA} is set to true when transition
\code{$Poll$} is fired, that is when \code{A} becomes true for the
first time. Transition \code{OK} is used to set \code{win} to true if
\code{A} is true for duration $D$ without interruption (otherwise its
timing constraint is resetted). Otherwise, if variable \code{win} is
still false after $D$ u.t., then transition \code{$Error$} is
fired. We use a priority between \code{$Error$} and \code{$OK$} to
disambiguate the behavior $D$ u.t. after \code{Poll} is fired.

\subsection{Absence patterns}
\label{realtimepatternsA}

Absence patterns are used to express that some condition should never
occur.
\smallskip

\fichepattern{\code{Absent} $A$ \code{after} $B$ \code{for interval} $I$}
{Predicate $A$ must never hold between $d_1$--$d_2$ u.t. after
the first occurrence of $B$.}
{This pattern is dual to \pop{Present} $A$ \pop{after} $B$ \pop{within} $I$
(it is not equivalent to its negation because, in both patterns, $B$ is not required to occur).
}
{\pop{absent} $\lbl{Open}_1 \vee \lbl{Open}_2$ \pop{after}  $\lbl{Close}_1 \vee \lbl{Close}_2$ \pop{for interval} $[0, 10]$}
{\neg B \W (B \wedge \ltlall_{I} \neg A)}
{\forall \s_1, \s_2, \s_3, \event \such (\s = \s_1 B \s_2 \event \s_3 \wedge B \notin \s_1 \wedge \Delta(\s_2) \in I) \imply \neg A(\event)}
{
We use the same observer as for \pop{Present} $A$ \pop{after} $B$ \pop{within} $I$,
but here \lbl{Error} is the expected behavior.
}
{\ltlexist B \imply \ltlexist \lbl{Error}}

\noindent Same as the explanation for \pop{Present} $A$ \pop{after}
$B$ \pop{within} $I$.  
\smallskip

\fichepattern{\code{Absent} $A$ \code{before} $B$ \code{for duration} $D$}
{No $A$ can occur less than $D$ u.t. before the first occurrence of~$B$. The pattern holds if there are no occurrence of $B$.}{}
{\pop{absent} \code{Open$_1$} \pop{before} \code{Close$_1$} \pop{for duration} $3$}
{\ltlexist B \imply (A \imply (\ltlall_{[0 , D]} \neg B)) \until B}
{\forall \s_1, \s_2, \s_3, \event \such (\s = \s_1 \event \s_2 B \s_3 \wedge B \notin \s_1\event\s_2 \wedge \Delta(\s_2) \leqslant D) \imply \neg A(\event)}
{
\begin{tikzpicture}[baseline,node distance=5ex and 9ex, label distance=-0.5ex]
         \node[vtransition, label=above:$E_1$, label=below:{\postact{\begin{tabular}[t]{rcl}bad &:=& true\\foundB &:=& false\end{tabular}}}] (t1) {} ;
          \node[vtransition, label=above:$E_2$, label=below:{\postact{foundB := true}}, left=22ex of t1] (t0) {} ;
          \node[place, right=of t1, tokens=1, label=above:{idle}] (p1) {} edge[pre,<->] (t1) ;
          \node[vtransition, right= of p1, label=above:Reset, label=below:{\postact{bad := false}}, label=right:{$[D, D]$}] (t2) {} edge[readarc] (p1) ;
          
    \end{tikzpicture}

}
{\ltlall \neg (\lbl{foundB} \wedge \lbl{bad})}

\noindent Variable \code{$foundB$} is set to true after each
occurrence of \code{$E_2$}. Conversely, we set the variables
\code{$bad$} to true and \code{$foundB$} to false at each occurrence
of \code{$E_1$}. Therefore \code{$foundB$} is true on every ``time
interval'' between an \code{$E_2$} and an \code{$E_1$}. We use
transition \code{$Reset$} to set \code{$bad$} to false if this
interval is longer than $D$. As a consequence, the pattern holds if we
cannot find an occurrence of \code{$E_2$} (\code{$foundB$} is true)
while \code{$bad$} is true.


\subsection{Response patterns}
\label{realtimepatternsR}

Response patterns are used to express ``cause--effect'' relationship,
such as the fact that an occurrence of a first kind of events must be
followed by an occurrence of a second kind of events.
\smallskip

\fichepattern{$A$ \code{leadsto} first $B$ \code{within} $I$}
{Every occurrence of $A$ must be followed by an occurrence of $B$
within  time interval $I$ (considering only the first occurrence of
$B$ after $A$).}{}
{\lbl{Button$_2$} \pop{leadsto first} \lbl{Open$_2$} \pop{within} $[0, 10]$}
{\ltlall(A \Rightarrow (\neg B) \until_{I}B)}
{
\forall \s_1, \s_2 \such (\s = \s_1 A \s_2) \Rightarrow
\exists \s_3,\s_4 \such \s_2 = \s_3 B \s_4 \wedge \Delta(\s_3)
\in I \wedge B \notin \s_3
}
{
\hspace{-9ex}
\begin{tikzpicture}[baseline,node distance=7ex and 8.5ex, label distance=-0.5ex]
          \node[vtransition, label=above:$E_1$, label=below:{\postact{\begin{tabular}[t]{rcl}foundA & := & true \\ bad & := & true\end{tabular}}}] (t1) {} ;
          \node[place, right=of t1, tokens=1] (p1) {} edge[pre,<->] (t1) ;
          \node[vtransition, right= of p1, label=above:Start, label=below:{\postact{bad := false}}, label=right:{$[d_1, d_1]$}] (t2) {} edge[readarc] (p1) ;
          \node[vtransition, right=20ex of t2, label=above:$E_2$, label=below:{\postact{foundA := false}}] (t0) {} ;
          \node[vtransition, right=16ex of t0, label=above:Error, label=below:{\preact{foundA}}, label=right:{$[d_2, d_2]$}] (t3) {} ;
 \draw[prio] (t2.south east) -- (t0.south west) ;
 \draw[prio] (t0.south west) -- (t3.south east) ;
\end{tikzpicture}
}
{(\ltlall \neg \lbl{Error}) \wedge (\ltlall \neg (B \wedge \lbl{bad}))}

\noindent After each occurrence of \code{$E_1$}, variables
\code{$foundA$} and \code{$bad$} are set to true and the transition
\code{$Start$} is enabled.  Variable \code{$bad$} is used to control
the beginning of the time interval. After each occurrence of
\code{$E_2$} variable \code{$foundA$} is set to false. Hence
\code{$Error$} is fired if there is an occurrence of \code{$E_1$} not
followed by an occurrence of \code{$E_2$} after $d_2$ u.t. We use
priorities to avoid errors when \code{$E_2$} occurs precisely at time
$d_1$ or $d_2$.
\smallskip

\fichepattern{$A$ \code{leadsto} first $B$ \code{within} $I$ \code{before} $R$}
{Before the first occurrence of $R$, each occurrence of $A$ is
  followed by a $B$---and these two events occur before $R$---in
 the time interval $I$. The pattern holds if $R$ never occur.}{}
{\lbl{Button$_2$} \pop{leadsto first} \lbl{Open$_2$} \pop{within} $[0, 10]$ \pop{before} \lbl{Shutdown}}
{\ltlexist R \imply (\ltlall (A \wedge \neg R \imply (\neg B \wedge \neg R) \until_{I} B \wedge \neg R) \until R}
{
\forall \s_1, \s_2, \s_3 \such (\s = \s_1 A \s_2 R \s_3 \wedge R \notin \s_1 A \s_2  \imply 
\exists \s_4,\s_5 \such \s_2 = \s_4 B \s_5 \wedge \Delta(\s_4) \in I \wedge B \notin \s_4
}
{
\hspace{-4ex}
\begin{tikzpicture}[baseline,node distance=7.0ex and 11ex, label distance=-0.5ex]
          \node[vtransition, label=left:$E_1$, label=below:{\postact{\begin{tabular}[t]{rcl} if $\neg$ foundR then foundA & := & true \\ bad & := & true\end{tabular}}}] (t1) {} ;
          \node[place, right=of t1, tokens=1] (p1) {} edge[pre,<->] (t1) ;
          \node[vtransition, right= of p1, label=above:Start, label=below:{\postact{bad := false}}, label=right:{$[d_1, d_1]$}] (t2) {} edge[readarc] (p1) ;
          \node[vtransition, right=20ex of t2, label=left:Error, label=below:{\preact{foundA}}, label=right:{$]d_2,\infty[$}] (t3) {} ;
          \node[vtransition, below=of t1, label=left:$E_2$, label=below:{\postact{if $\neg$ foundR then foundA := false}}] (t0) {} ;
          \node[vtransition, below=of t2, label=left:$E_3$, label=below:{\postact{foundR=true}}] (t4) {} ;
\end{tikzpicture}
}
{\ltlexist R \imply (\ltlall \neg \lbl{Error} \wedge \ltlall \neg (B \wedge \lbl{bad}))}

\noindent Same explanation than for the previous case, but we only
take into account transitions \code{$E_1$} and \code{$E_2$}
occurring before \code{$E_3$}.
\smallskip

\fichepattern{$A$ \code{leadsto first} $B$ \code{within} $I$ \code{after} $R$}
{Same than with the pattern ``$A$ \pop{leadsto first} $B$ \pop{within} $I$''
  but only considering occurrences of $A$ after the first $R$.}{}
{\lbl{Button$_2$} \pop{leadsto first} \lbl{Open$_2$} \pop{within} $[0, 10]$ \pop{after} \lbl{Shutdown}}
{\ltlall (R \imply (\ltlall (A \imply (\neg B) \until_{I} B)))}
{\forall \s_1, \s_2 \such (\s = \s_1 R \s_2 A \s_3 \wedge R \notin \s_1) \imply
  \exists \s_4,\s_5 \such \s_3 = \s_4 B \s_5 \wedge \Delta(\s_4) \in I \wedge B \notin \s_4
}
{It is similar to the observer of the pattern $A$ \code{leadsto} first $B$ \code{within} $I$ \code{before} $R$ . We should just replace \lbl{$\neg$foundR} in transition \lbl{$E_1$} and \lbl{$E_2$} by \lbl{foundR}.}
{\ltlexist R \imply  (\ltlall \neg \lbl{Error} \wedge \ltlall \neg (B \wedge \lbl{bad}))}

\noindent Same explanation than in the previous case, but we only take
into account transitions \code{$E_1$} and \code{$E_2$} occurring after
an \code{$E_3$}.

\subsection{Composite Patterns}
\label{realtimepatternsC}

Patterns can be easily combined together using the usual boolean
connectives.  For example, the pattern ``$P_1$ \pop{and} $P_2$'' holds
for all the traces where $P_1$ and $P_2$ both hold. To check a
composed pattern, we use a combination of the respective observers, as
well as a combination of the respective LTL formulas. For instance, if
$(T_1, \phi_1)$ and $(T_2, \phi_2)$ are the observers and LTL formulas
corresponding to the patterns $P_1$ and $P_2$, then the composite
pattern $P_1$ \pop{and} $P_2$ is checked using the LTL formula $\phi_1
\wedge \phi_2$. Similarly, if we check the LTL formula $\phi_1
\Rightarrow \phi_2$ (implication) then we obtain a composite pattern
$P_1 \multimap P_2$ that is satisfied by systems $T$ such that, for
all traces of $T$, the pattern $P_2$ holds whenever $P_1$ holds.

\section{Use Cases and Experimental Results}
\label{sec:use-cases-exper}

In this section, we report on three experiments that have been
performed using an extension of a Fiacre compiler that automatically
compose a system with the necessary observers. In case the system does
not meet its specification, we obtain a counter-example that can be
converted into a timed sequence of events exhibiting a problematic
scenario. This sequence can be played back using \emph{play} and
\emph{nd}, two Time Petri Nets animators provided with Tina.

\paragraph*{Avionic Protocol and AADL.} Our first example is a network
avionic protocol (NPL) which includes several functions allowing the
pilot and ground stations to receive and send information relative to
the plane: weather, speed, \dots AADL has been used to model the
dynamic architecture for this demonstrator~\cite{erts}. The AADL model
includes several threads that exchange information through shared
memory data and amounts to about $8$ diagrams and $800$ lines of code
(using AADL textual syntax). The AADL code specifies both the hardware
and software architecture of the system and defines the real time
properties of threads, like for instance their dispatch protocol
(periodic or sporadic) or their periods.


We used the AADL2Fiacre plug-in of Topcased to check properties on the
NPL specification. The Fiacre model obtained after transformation
takes into account the complete behavior described in the AADL model
but also the whole language execution model, meaning that our
interpretation takes fully into account the scheduling semantics as
specified in the AADL standard. The abstract state space for the TTS
generated from Fiacre has about 120\,000 states and 180\,000
transitions and can be generated in less than 12\+s on a typical
development computer (Intel dual-core processor at 2\+GHz with 2\+Gb
of RAM). On examples of this size, our model checker is able to prove
formal properties in a few seconds. We checked a set of 22
requirements that were given together with the description of the
system, all expressed using a natural language description and, in one
case, a scenario based on a UML sequence diagram. Of these 22
requirements, 18 where instances of ``untimed patterns'', such as
checking the absence of deadlock or that threads are resettable. The
four remaining requirements where ``response patterns'' of the kind
$A$ \code{leadsto} first $B$ \code{within} $[0, d]$. Using patterns,
we were able to check the 22 patterns in less than 5\+min.


\paragraph*{Service Oriented Applications.} We consider models
obtained from the composition of services expressed using a timed
extension of BPEL, the Business Process Execution Language. Our
example models a scenario from the health-care domain related to
patient handling during a medical examination. The scenario involves
three entities, each one managed by a service: a Clinic Service (CS);
a Medical Analysis Service (MAS); and a Pharmacy Service (PS). When a
patient arrives in clinic, a doctor should check with the MCS whether
its social security number is valid. If so, the doctor may order some
medical analyses from the MAS and, after analyzing the results, he can
order drugs through the PS. Timing constraints can be added to this
scenario by associating a duration to each activity of the workflow
and a delay to each service invocation.

We use our patterns to express different requirements on this
system. An example involving the absence pattern is that we cannot
have two medical analyses for a patient in less than 10 days (240
hours): \code{absent} MAS.medicalAnalysis \code{after}
MAS.medicalAnalysis \code{for interval} $]0, 240]$. A more complicated
example of requirement is to impose that if a doctor does not cancel a
drug order within 6 hours, then it should not cancel drugs for another
48 hours. This requirement can be expressed using the composition of
two absence patterns (see Sect.~\ref{realtimepatternsC}):
\[
\begin{array}[c]{l}
  \text{(\code{absent} MCS.drugsChanging \code{after} MCS.drugsAsking
    \code{for interval} $[0; 6]$)}\\
  \quad \multimap \text{(\code{absent}
    MCS.drugsChanging \code{after} MCS.drugsAsking \code{for interval}
    $[0; 54]$).}
\end{array}
\]
Finally, using the notation S.init and S.end to refer to a start
(resp. end) event in the service S, we can express that drugs must be
delivered within 48 hours of the medical examination start: MCS.init
\code{leadsto} PS.sendDrugsOrder \code{within} $[0; 48]$.

The complete scenario is given in~\cite{bpel}, where we describe a
transformation tool chain from Timed BPEL processes to Fiacre. For a
more complex version of the health care scenario, with seven different
services and more concurrent activities, the state graph for the TTS
generated from Fiacre is quite small, with only 886 states and 2476
transitions. The generation of the Fiacre specification and its
corresponding state space takes less than a second. For examples of
this size, the verification time for checking a requirement is
negligible (half a second).



\paragraph*{Transportation Systems.} Our final example is an automated
railcar system, taken from~\cite{tap}, that was directly modeled using
Fiacre. The system is composed of four terminals connected by rail
tracks in a cyclic network. Several railcars, operated from a central
control center, are available to transport passengers between
terminals. When a car approaches its destination, it sends a request
to signal its arrival to the terminal. This system has several
real-time constraints: the terminal must be ready to accommodate an
incoming car in 5\+s; a car arriving in a terminal leaves its door
open for exactly 10\+s; passengers entering a car have 5\+s to choose
their destination; etc. There are three key requirements:

\noindent\quad({P1}) when a passenger arrives in a terminal, a car must
be ready to transport him within 15\+s. This property can be expressed
with a response pattern, where \code{Passenger/sndReq} is the state
where the passenger requests a car and \code{Car/ackTerm} is the state
where it is served:
\[ \text{\code{Passenger/sendReq} \pop{leadsto} \code{Car/ackTerm}
  \pop{within}} [0,15] \]
\quad({P2}) When the car starts moving, the door must be closed:
\[ \text{\pop{present} \code{CarDoor/closeDoor} \pop{after}
\code{CarHandler/moving} \pop{within}} [0,10] \]
\quad({P3}) When a passenger select a destination (in the car), a
signal should stay illuminated until the car is arrived: 
\[ \text{\pop{absent} \code{Terminal/buttonOff} \pop{before}
  \code{Control/ackTerm} \pop{for duration}} 10 \]

We can prove that these three patterns are valid on our Fiacre
model. Concerning the performances, we are able to generate the
complete state space of the railcar system in 310\+ms, using 400\+kB
of memory. This gives an upper-bound to the complexity of checking
simple (untimed) reachability properties on the system, like for
instance the absence of deadlocks. The three patterns can all be
checked in under 1.5\+s. For instance, we observed that checking
property ({P1}) is not more complex than exploring the complete
system: the property is checked in 450\+ms, using 780\+kB of
memory. Also, this is roughly the same complexity than checking the
corresponding untimed requirement in LTL that is:
$\ltlall$\,(\code{Passenger/sendReq}~$\imply
\ltlexist$\code{Control/ackTerm}).


\paragraph*{Conclusion.} In other benchmarks, we have often found that
the complexity of checking timed patterns is in the same order of
magnitude than checking their untimed temporal logic equivalent. An
exception to this observation is when the temporal values used in the
patterns are far different from those found in the system; for example
if checking a periodic system, with a period in the order of the
milliseconds, against a requirement using an interval in the order of
the minutes. More results on the complexity of our approach can be
found in~\cite{VRTSPTTS}.  These experimentation, while still modest
in size, gives a good appraisal of the use of formal verification
techniques for real industrial software.

These experimental results are very encouraging. In particular, we can
realistically envisage that system engineers could evaluate different
design choices in a very short time cycle and test the safety of their
solutions at each iteration.

\section{Related Work and Contributions}
\label{sec:related-work}

We base our approach on the original catalog of specification patterns
defined by Dwyer~\cite{ppsfsv1999}. This work essentially study the
expressiveness of their approach and define patterns using different
logical framework (LTL, CTL, Quantified Regular Expressions, etc.).
As a consequence, they do not need to consider the problem of checking
requirements as they can readily rely on existing
model checkers. Their patterns language is still supported, with
several tools, an online repository of examples~\cite{ksu} and the
definition of the Bandera Specification Language~\cite{alfecpds} that
provides a structured-English language front-end. A recent study by
Bianculli et al.~\cite{SPRICSSBA} show the relevance of this
pattern-based approach in an industrial context.

Some works consider the extension of patterns with hard real-time
constraints. Konrad et al.~\cite{RSP} extend the patterns language
with time constraints and give a mapping from timed pattern to TCTL
and MTL. Nonetheless, they do not consider the complexity of the
verification problem (the implementability of their approach). Another
related work is~\cite{PTPS}, where the authors define observers based
on Timed Automata for each pattern. However, they consider a less
expressive set of patterns (without the lasting modifier) and they
have not integrated their language inside a tool chain or proved the
correctness of their observers. By contrast, we have defined a formal
framework that has been used to prove the correctness of some of our
observers~\cite{VRTSPTTS}. Work is currently under way to mechanize
these proofs using the Coq interactive theorem prover. 

Our patterns can be viewed as a subset of the Contracts Specification
Language (CSL), defined in the context of the SPEEDS
project~\cite{csl}, which is intended as a pragmatic proposal for
specifying contract assertions on HRC models. While the semantics for
HRC is based on hybrid automata, the only automatic verification tools
available for CSL use a discrete time model. Therefore, our
verification tool chain provides a partial implementation for CSL (the
part concerned by timing constraints) for a dense time model. This is
an important result since more conception errors can be captured using
a dense rather than a discrete time model.


Compared to these related works, we make several contributions. We
extend the specification patterns language of Dwyer et al. with two
modifiers for real-time constraints. We also address the problem of
checking the validity of a pattern on a real-time system using
model-based techniques: our verification approach is based on a set of
observers, that are described in Sect.~\ref{sec4:RPC}. Using this
approach, we reduce the problem of checking real-time properties to
the problem of checking simpler LTL properties on the composition of
the system with an observer. Another contribution is the definition of
a formal framework to prove that observers are correct and
non-intrusive, meaning that they do not affect the system under
observation. This framework is useful for proving the soundness of
optimization. Due to space limitations, we concentrate on the
definition of the patterns and their semantics in this paper, while
most of the theoretical results are presented in a companion research
report~\cite{VRTSPTTS}.  Finally, concerning tooling, we offer an
EMF-based meta-model for our specification patterns that allow its
integration within a model-driven engineering development: our work is
integrated in a complete verification tool chain for the Fiacre
modelling language and can therefore be used in conjunction with
Topcased~\cite{ttptosfcasd2006}, an Eclipse based toolkit for system
engineering.

\section{Conclusion and Perspectives}

We define a set of high-level specification patterns for expressing
requirements on systems with hard real-time constraints. Our approach
eliminates the need to rely on model checking algorithms for timed
extensions of temporal logics that---when decidable---are very complex
and time-consuming. While we have concentrated our attention on
model checking---and although we only provide an implementation for
TTS models---we believe our notation is interesting in its own right
and can be reused in different contexts.

There are several directions for future works. We plan to define a
compositional patterns inspired by the ``denotational interpretation''
used in the definition of patterns. The idea is to define a
lower-level pattern language, with more composition operators, that is
amenable to an automatic translation into observers (and therefore can
dispose with the need to manually prove the correctness of our
interpretation). In parallel, we plan to define a new modelling
language for observers---adapted from the TTS framework---together
with specific optimization techniques and easier soundness
proofs. This language, which has nearly reached completion, would be
used as a new target for implementing patterns verification.




 

 
\end{document}